\documentclass[onecolumn,structabstract]{aa}
\usepackage[latin1]{inputenc}
\usepackage[dvips]{graphicx}
\usepackage{amssymb}
\usepackage{amsmath}
\usepackage{txfonts}
\usepackage[comma,authoryear]{natbib}
\newcommand{\eg}{\textit{e.g.}}
\newcommand{\ie}{\textit{i.e.}}

\begin{document}

\title{Stable higher order finite-difference schemes for stellar pulsation
       calculations}
\author{D. R. Reese\inst{1,2}}
\institute{Institut d'Astrophysique et Géophysique de l'Université de Liège,
           Allée du 6 Août 17, 4000 Liège, Belgium \\
           \email{daniel.reese@ulg.ac.be}
           \and
           Observatoire de Paris, LESIA, CNRS UMR 8109, 92195 Meudon, France}

\abstract
{Calculating stellar pulsations requires a sufficient accuracy to match the
quality of the observations.  Many current pulsation codes apply a second order
finite-difference scheme, combined with Richardson extrapolation to reach fourth
order accuracy on eigenfunctions.  Although this is a simple and robust
approach, a number of drawbacks exist thus making fourth order schemes
desirable.  A robust and simple finite-difference scheme, which can easily be
implemented in either 1D or 2D stellar pulsation codes is therefore required.}
{One of the difficulties in setting up higher order finite-difference schemes
for stellar pulsations is the so-called mesh-drift instability. Current ways of
dealing with this defect include introducing artificial viscosity or applying a
staggered grids approach.  However these remedies are not well-suited to
eigenvalue problems, especially those involving non-dissipative systems, because
they unduly change the spectrum of the operator, introduce supplementary free
parameters, or lead to complications when applying boundary conditions.}
{We propose here a new method, inspired from the staggered grids strategy, which
removes this instability while bypassing the above difficulties.  Furthermore,
this approach lends itself to superconvergence, a process in which the accuracy
of the finite differences is boosted by one order.}
{This new approach is successfully applied to stellar pulsation calculations,
and is shown to be accurate, flexible with respect to the underlying grid,
and able to remove mesh-drift.}
{Although specifically designed for stellar pulsation calculations, this method
can easily be applied to many other physical or mathematical problems.}


\maketitle

\section{Introduction}

Numerically calculating pulsation modes in stellar models typically involves
solving a differential system for which the coefficients change by several
orders of magnitude, and in which the underlying grid can be highly uneven. 
Currently, many stellar pulsation codes use a second order finite-difference
scheme combined with Richardson extrapolation in order to achieve sufficient
accuracy on the frequencies \citep{Richardson1927, Moya2008}.  Although this
approach is robust and simple, it has some disadvantages. Firstly, only the
frequencies are accurate to fourth order; the eigenfunctions remain second order
accurate. Secondly, Richardson extrapolation requires calculating the pulsation
modes for two different resolutions, typically with $N$ and $N/2$ points.  These
solutions then need to be matched correctly before applying the extrapolation. 
Although this is straightforward in most cases, it may be more challenging in
red giant stars. Indeed, non-radial modes in such stars have a mixed p- and
g-mode behaviour, thus resulting in a high density of nodes in the inner regions
of the star \citep[\eg\ ][]{Dupret2009}.  Special care is needed to correctly
match the modes, which furthermore need to be fully resolved at both
resolutions.  A third difficulty, which is illustrated in
Section~\ref{sect:Richardson}, is that Richardson extrapolation is only strictly
valid when the additional points from the denser grid lie at the midpoints of
the coarser grid.

A way to a circumvent these difficulties is to apply a fourth or higher order
scheme.  This is the approach taken in the OSCROX \citep{Roxburgh2008},
CAFein \citep{Valsecchi2013}, PULSE \citep{Brassard2008}  and LOSC
\citep{Scuflaire2008} oscillations codes.  When devising the stellar oscillation
code TOP \citep[Two-dimensional Oscillation Program][]{Reese2006,Reese2009}, a
fourth or higher order approach was also searched for.  Given that this code
uses an algebraic rather than a shooting method for finding the eigensolutions,
applying a fourth order Runge-Kutta method, such as what is done in OSCROX
and CAFein, is inappropriate.  \citet{Gautschy1990} and
\citet{Takata2004}, upon which the CAFein code is built, use predictor-corrector
schemes to integrate the pulsation equations, which were reformulated through
the Riccati method.  Such schemes can be set up to fourth or higher order
accuracy, but much like the Runge-Kutta integrator, are also unsuitable for an
algebraic approach.  Furthermore, TOP is designed to handle pulsation modes in
rapidly rotating stars, a 2D problem, thus requiring a simple and robust
scheme.  The PULSE code uses a finite element approach
\citep{Brassard1992}, which although accurate and stable, leads to a higher
complexity in the program due to the need to integrate the products of
equilibrium quantities and basis functions over each element.  A
finite-difference approach is simpler, since the discretised form of the
differential equations only involve coefficients which are the products of
equilibrium quantities and finite-difference coefficients. Accordingly, the
approach implemented in the LOSC code is closest to these requirements since it
is a fourth order finite-difference scheme \citep{Scuflaire2008}.  Nonetheless,
the main drawback with this scheme is that it requires calculating radial
derivatives of equilibrium quantities, which can be an additional source of
error, especially in evolved models with steep compositional, and hence density,
gradients.  Therefore, a simpler fourth or higher order finite-different scheme
was searched for when developing TOP.

A first, and somewhat naive, form of higher order finite differences was
initially implemented in TOP \citep{Reese2009}.  However, a numerical
instability known as mesh-drift appeared, thereby leading to oscillatory
behaviour and spurious modes.  Prior to this, \citet{Dupret2001} also reported
on similar difficulties when setting up the non-adiabatic pulsation code called
MAD.  The goal of the present paper is therefore to search for some of the
causes behind mesh drift and to propose a remedy.  In the following section, the
problem of mesh drift is described, illustrated and characterised. 
Section~\ref{sect:alternate} then proposes a new method for dealing with mesh
drift and compares it with other approaches. The following section shows how
this new approach can be enhanced through ``superconvergence'', which boosts the
order of the finite differences by one.  Section~\ref{sect:application}
describes and evaluates 1D and 2D stellar pulsation calculations based on this
new approach.  Finally, the paper ends with a short conclusion.

\section{Mesh drift}

It has been known for many years that finite-difference schemes can suffer from
mesh-drift instability. This instability typically involves the decoupling of
even and odd grid points in the differences equations \citep[\eg][]{Press2007}. 
Depending on the differential problem which is being solved, this can lead to
solutions with highly oscillatory behaviour or spurious solutions.

\subsection{Examples of mesh-drift instability}
\label{sect:examples}

In order to illustrate some of the difficulties caused by mesh drift, we propose
to numerically solve the following trivial eigenvalue problem over the interval
$[0,1]$:
\begin{equation}
\label{eq:example}
v' = -\omega u, \qquad u' = \omega v, \qquad u(0) = u(1) = 0.
\end{equation}
where $\omega$ is the eigenvalue and primes denote first order derivatives with
respect to the variable $x$. The solutions are $u(x) = A \sin(\omega x)$, $v(x) =
A \cos(\omega x)$, where $\omega = k\pi$, $k$ is an integer, and $A$ an arbitrary
non-zero constant.

\begin{figure}[htbp]
\begin{center}
\includegraphics[width=0.8\textwidth]{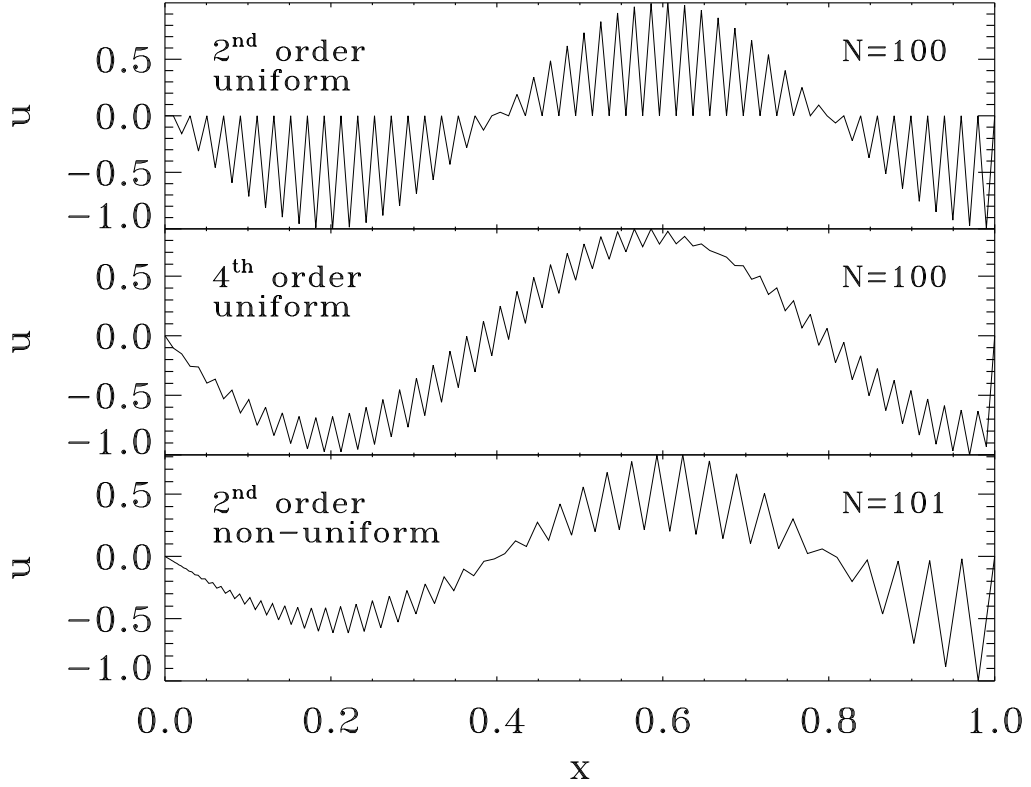}
\end{center}
\caption{Spurious numerical solutions to Eq.~(\ref{eq:example}) caused by mesh
drift, using a 2$^{\mathrm{nd}}$ order finite-difference scheme on a uniform
grid (\textit{upper panel}), a 4$^{\mathrm{th}}$ order scheme on a uniform grid
(\textit{middle panel}), and a 2$^{\mathrm{nd}}$ order scheme on a non-uniform
grid. The number of grid points, $N$, is indicated in the upper right corner of
each panel.\label{fig:mesh_drift}}
\end{figure}

In the first case, we use a uniform grid and a second order scheme defined over
a three-point window.  On the end points, we replace the above formula by a
first order scheme defined on a two-point window so as to avoid introducing
supplementary bands in the matrix representation of the derivation operator
which takes on a banded form.  Avoiding supplementary bands can be important for
reducing the numerical cost of large scale computations, where the number of
bands in the total system depends critically on that of the derivation
operator.  The boundary conditions are imposed by replacing two of the $2N$
difference equations.  Hence, the resultant difference equations are:
\begin{equation}
\begin{array}{rclrcl}
-\omega u_i &=& \displaystyle \frac{v_{i+1} - v_{i-1}}{2\Delta x}, \qquad &
 \omega v_i &=& \displaystyle \frac{u_{i+1} - u_{i-1}}{2\Delta x}, \qquad 2 \leq i \leq N-1, \\
-\omega u_N &=& \displaystyle \frac{v_N - v_{N-1}}{\Delta x}, &
 \omega v_1 &=& \displaystyle \frac{u_2 - u_1}{\Delta x}, \\
        u_1 &=& 0,  &  u_N &=& 0,
\end{array}
\end{equation}
where $N$ is the number of grid points, and $\Delta x = x_{i+1} - x_i$. Although
the correct solutions are well recovered, supplementary spurious solutions
appear such as the one in the upper panel of Fig.~\ref{fig:mesh_drift},
with an eigenvalue around $2.5\pi$.

One may then try a fourth order scheme using windows with 5 grid points
$(x_{i-2}, ... x_{i+2})$ with reduced order schemes near the endpoints and
boundary conditions which replace 2 of the difference equations, as was done
above.  The middle panel of Fig.~\ref{fig:mesh_drift} illustrates a similar
spurious solution.

Finally, the last panel shows the results for a second order scheme using
3-point windows (except at the endpoints) and using a non-uniform grid:
\begin{equation}
x_i = \frac{(i-1)^2}{(N-1)^2}, \qquad 1 \leq i \leq N.
\label{eq:grid_non_uniform}
\end{equation}
Furthermore, we changed the number of grid points from $N=100$ to
$N=101$. The finite-difference weights involved were deduced from
\citet{Fornberg1988}. Once more, similar spurious solutions appear, such as the
one illustrated in the lower panel of Fig.~\ref{fig:mesh_drift}.  These last two
examples show that mesh drift is, in fact, a more general problem than simply a
decoupling between even and odd grid points, given that all the points are
coupled in these latter two schemes.  Also, the last example shows that the
problem occurs regardless of whether $N$ is even or odd.  This raises the
questions as to when mesh drift occurs and how to remove it.

\subsection{Characterising mesh drift}
\label{sect:characterising}

In the first example above, mesh drift occurred because even and odd points were
decoupled.  Another way of looking at this is that the numerical derivative of a
function, $f$, defined such that $f(x) = a$ for even points and $f(x) = b$ for
odd points, where $a$ and $b$ are two distinct constants, will be zero (except
at the end points where a different scheme is used).  Hence, one may generalise
this and look for functions which numerically satisfy the relation $f' = 0$ for
other schemes.

In order to avoid only having a constant solution, we impose $f' = 0$ only over
the interval which allows full windows for the different schemes above.  Hence,
for a 3-point second order scheme, we will impose:
\begin{equation}
a^i_{-1} f_{i-1} + a^i_0 f_i + a^i_1 f_{i+1} = 0, \qquad 2 \leq i \leq N-1,
\end{equation}
where $a^i_j$ denotes finite-difference weights at point $i$ (in the case of a
uniform grid, $a^i_j$ will only depend on j).  For a 5-point fourth order
scheme, we have:
\begin{equation}
a^i_{-2} f_{i-2} + a^i_{-1} f_{i-1} + a^i_0 f_i + a^i_1 f_{i+1} + a^i_2 f_{i+2} = 0,
\qquad 3 \leq i \leq N-2.
\end{equation}
For the second order schemes, the resultant null space will be of dimension
$D=2$ since there are  $N-2$ equations for $N$ unknowns.  Likewise, the null
space for 4$^\mathrm{th}$ order schemes will be of dimension $D=4$.  In both
cases, the null space will include constant functions, which are the correct
solutions to the problem.

An orthogonal basis\footnote{Orthogonal with respect to the dot product $f \cdot
g = \sum_i f_i g_i$.  This implies that that the basis elements will be
normalised as follows: $\sum_i f_i^2 = 1$.}, $\tilde{\mathcal{B}} =
(\tilde{b}^1, \tilde{b}^2, ... \tilde{b}^D)$, for the null space of the above
system can be extracted by calculating, for example, a singular value
decomposition of its representative matrix (made up of the coefficients
$a^i_j$).  However, to go one step further, it is helpful to find a new basis
which brings out the most oscillatory solutions.  We therefore introduce the
following quadratic form which measures how oscillatory a given function,
$\left(g_i\right)_{i = 1 \dots N}$, is:
\begin{equation}
J(g) = \sum_{i=2}^{N-1} \left( \frac{g_{i+1} - g_i}{x_{i+1} - x_i} - \frac{g_i - g_{i-1}}{x_i - x_{i-1}}\right)^2,
\label{eq:oscillatoriness}
\end{equation}
and search for normalised elements which maximise this measure. It is
straightforward to see that $J=0$ only for elements with a constant slope (\ie\
$g_i = ax_i+b$, where $a$ and $b$ are constants).  To make sure we find solutions
within the null space, we decompose $g$ over the $\tilde{\mathcal{B}}$ basis,
\ie\ $g = \sum_{j=1}^D \mu_j \tilde{b}^j$, and re-express the quadratic form in
terms of the coordinates $(\mu_j)_{j = 1 \dots D}$:
\begin{equation}
J\left((\mu_j)_{j = 1 \dots D}\right) = \sum_{i=2}^{N-1} \left\{
\sum_{j=1}^{D} \mu_j \left(  \frac{\tilde{b}_{i+1}^j - \tilde{b}_i^j}{x_{i+1} - x_i}
- \frac{\tilde{b}_i^j - \tilde{b}_{i-1}^j}{x_i - x_{i-1}} \right) \right\}^2
= \sum_{j=1}^{D} \sum_{k=1}^D  \mu_j \mu_k \left\{ \sum_{i=2}^{N-1}
\left(  \frac{\tilde{b}_{i+1}^j - \tilde{b}_i^j}{x_{i+1} - x_i}
- \frac{\tilde{b}_i^j - \tilde{b}_{i-1}^j}{x_i - x_{i-1}} \right)
\left(  \frac{\tilde{b}_{i+1}^k - \tilde{b}_i^k}{x_{i+1} - x_i}
- \frac{\tilde{b}_i^k - \tilde{b}_{i-1}^k}{x_i - x_{i-1}} \right) \right\}
\label{eq:oscillatoriness_reduced}
\end{equation}
By virtue of the Rayleigh-Ritz theorem, the normalised vectors which maximise
and minimise $J$ are the eigenvectors associated with the maximal and minimal
eigenvalues of $J$.  Moreover, the set of eigenvectors of $J$ can be
expressed as a new orthogonal basis, $\mathcal{B} = (b^1, b^2 ... b^D)$, of the
null space\footnote{Initially, this basis will be expressed in terms of the
reduced coordinates $(\mu_j)$ but can easily be re-expressed over the initial
grid using the elements of $\tilde{\mathcal{B}}$. Orthogonality of the resultant
basis follows from that of the basis expressed in terms of the reduced
coordinates.}.

In order to illustrate how the above procedure works, we apply it  to
the 2$^{\mathrm{nd}}$ order scheme defined over a uniform grid.  This leads to
the following set of equations:
\begin{equation}
\frac{f_{i+1} - f_{i-1}}{2\Delta_x} = 0, \qquad 2 \leq i \leq N-1.
\label{eq:uniform_2nd}
\end{equation}
A singular value decomposition of the representative matrix yields an
orthogonal basis of the null space, for example:
\begin{equation}
\begin{array}{rcl}
\tilde{b}^1_i &=& \left\{
\begin{array}{lr}
\displaystyle \frac{\cos \theta - \sin \theta}{\sqrt{N}},\qquad i = 1, 3, ... N-1 \\
\displaystyle \frac{\cos \theta + \sin \theta}{\sqrt{N}},\qquad i = 2, 4, ... N
\end{array} \right. \\
\tilde{b}^2_i &=& \left\{
\begin{array}{lr}
\displaystyle \frac{-\sin \theta - \cos \theta}{\sqrt{N}},\qquad i = 1, 3, ... N-1 \\
\displaystyle \frac{-\sin \theta + \cos \theta}{\sqrt{N}},\qquad i = 2, 4, ... N
\end{array} \right.
\end{array}
\end{equation}
where $\theta$ is an arbitrary number and where we have assumed, for
simplicity, that $N$ is even.  One can easily verify that the above functions
satisfy Eq.~(\ref{eq:uniform_2nd}) and form an orthogonal basis. However, unless
$\theta$ is a multiple of $\pi/2$, both basis functions contain both an
oscillatory and a constant component.  Nonetheless, what we want is a basis in
which one of the two solutions is the most oscillatory, and the other, the least
oscillatory, \textit{i.e.} constant.  This is where the second step of the
procedure intervenes.  Using Eq.~(\ref{eq:oscillatoriness_reduced}), we begin by
expressing the quadratic form $J$ in terms of the reduced coordinates $(\mu_1,
\mu_2)$:
\begin{equation}
J(\mu_1,\mu_2) = 
\left[ \begin{array}{cc} \mu_1 & \mu_2 \end{array} \right]
\left[\begin{array}{cc}
\Lambda \sin^2 \theta & \Lambda \sin\theta \cos\theta \\
\Lambda \sin\theta \cos\theta & \Lambda \cos^2 \theta
\end{array} \right]
\left[ \begin{array}{c} \mu_1 \\ \mu_2 \end{array} \right],
\end{equation}
where $\Lambda = \frac{16(N-2)}{N(\Delta x)^2}$.  The eigenvectors of
$J$ are $(\sin\theta,\cos\theta)$, associated with the eigenvalue $\Lambda$, and
$(\cos\theta,-\sin\theta)$, associated with the eigenvalue $0$.  This leads to
the following orthogonal basis:
\begin{equation}
\label{eq:null_space_analytical}
\begin{array}{rclcl}
b^1_i &=& (\sin\theta) \tilde{b}^1_i + (\cos\theta) \tilde{b}^2_i &=& 
\left\{ \begin{array}{lr}
\displaystyle - \frac{1}{\sqrt{N}},\qquad i = 1, 3, ... N-1 \\
\displaystyle   \frac{1}{\sqrt{N}},\qquad i = 2, 4, ... N
\end{array} \right. \\
b^2_i &=& (\cos\theta) \tilde{b}^1_i - (\sin\theta) \tilde{b}^2_i &=& \displaystyle \frac{1}{\sqrt{N}}.
\end{array}
\end{equation}
Thanks to this second step, the oscillatory and constant components have
been separated.

\begin{figure}[htbp]
\begin{center}
\includegraphics[width=0.8\textwidth]{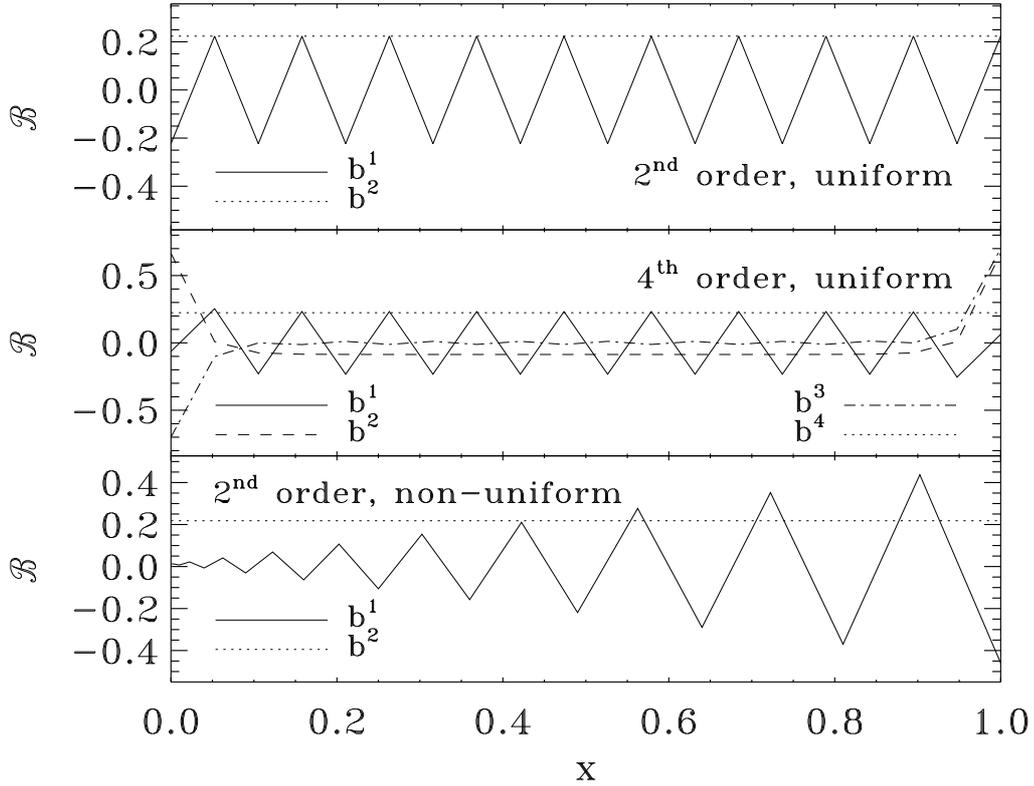}
\end{center}
\caption{Null space basis functions for the finite-difference schemes used in
Fig.~\ref{fig:mesh_drift} (except that the resolution has been reduced for the
sake of legibility).  The functions $(b^1, b^2 ... b^D)$ are ordered such that
the most oscillatory function comes first.\label{fig:null_space}}
\end{figure}

Figure~\ref{fig:null_space} illustrates the basis $\mathcal{B}$ obtained
numerically for the three cases considered in the previous section (except that
the number of grid points has been reduced to $N=20$, $N=20$,
and $N=21$, respectively, for the sake of legibility).  In each panel,
the most and least oscillatory solutions, $b^1$ and $b^D$, are shown as a
continuous and dotted line, respectively. The latter is simply a constant
function, as was expected.  The functions $b^1$ do turn out to be quite
oscillatory.  The basis functions in the top panel correspond, in fact,
to the analytical solutions given in Eq.~(\ref{eq:null_space_analytical}).

One may argue that enforcing $f' = 0$ all the way to the endpoints instead of
stopping before, as was done above, would filter out such oscillatory
solutions.  Although this may be true, it still means that \textit{the
oscillatory behaviour is being suppressed by the treatment near the endpoints
rather than by the finite-difference scheme throughout most of the domain},
implying that mesh drift is still likely to occur, as was observed in the
previous section.  Ideally, the basis functions should be nearly constant in the
centre and only deviate near the edges where $f'=0$ is not enforced.

\begin{figure}
\begin{minipage}{0.6\textwidth}
\includegraphics[width=\textwidth]{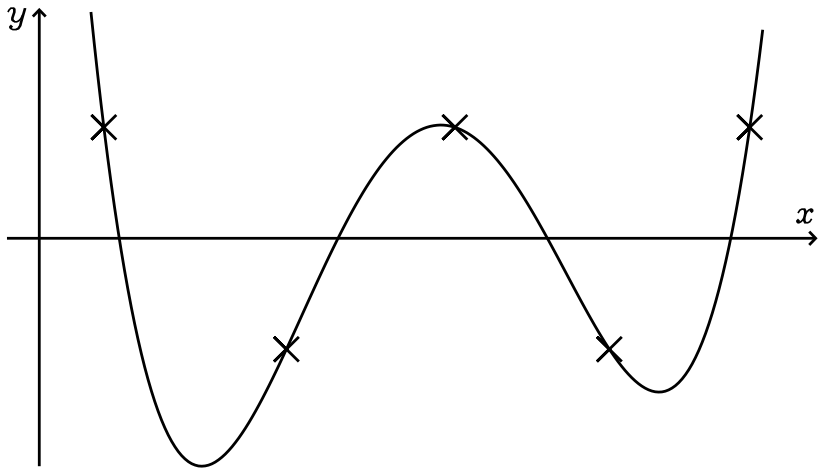}
\end{minipage} \hfill
\begin{minipage}{0.37\textwidth}
\caption{An oscillatory function (represented by the ``$\times$'') and
its interpolating polynomial (represented by the continuous line), defined over
a slightly non-uniform 5-point window.\label{fig:oscillatory}}
\end{minipage}
\end{figure}

The question therefore arises as to why such oscillatory functions are present,
even when even and odd points are coupled?  In order to provide a rough answer
to this question, it is helpful to look at a single window of grid points.
In Fig.~\ref{fig:oscillatory} we consider a 5 point window, with a
slightly non-uniform spacing between consecutive grid points.  We then define a
function $f$ such that it alternates between the values $1$ and $-1$ over this
window.  Calculating the function's derivative at a specified point, $x_0$,with a
finite-difference scheme amounts to finding the interpolation polynomial which
coincides with $f$ at the grid points, and then calculating the polynomial's
derivative at $x_0$.  Such a polynomial is illustrated in
Fig.~\ref{fig:oscillatory}.  As can be seen, the middle grid point is near a
local maximum.  Hence, the derivative of the polynomial is close to zero at that
point, meaning that the finite-difference scheme will give a result close to
zero for the numerical derivative, even though $f$ is oscillatory. 
Consequently, in order to avoid having small derivatives for oscillatory
functions, one needs to avoid calculating the derivative on or near one of the
local minima or maxima.  One could therefore chose one of the
endpoints, or an intermediate point away from local extrema.  In what
follows, we explore the latter possibility by considering windows with
an even number of points and calculating the first derivative at a point midway
between the two middle grid points.

\section{Finite differences enforced on an alternate grid}
\label{sect:alternate}

If we consider, once more, our initial problem (Eq.~(\ref{eq:example})), we can
set up the following, frequently used, second order scheme using 2-point
windows:
\begin{equation}
\begin{array}{rcl}
-\displaystyle \omega \frac{u_i + u_{i+1}}{2} &=&
 \displaystyle \frac{v_{i+1} - v_i}{x_{i+1} - x_i}, \qquad 1 \leq i \in \leq N-1, \\
 \displaystyle \omega \frac{v_i + v_{i+1}}{2} &=&
 \displaystyle \frac{u_{i+1} - u_i}{x_{i+1} - x_i}, \qquad 1 \leq i \in \leq N-1, \\
 u_1 &=& 0,  \qquad  u_N = 0.
\end{array}
\label{eq:two_points}
\end{equation}
Although the unknown functions, $(u_i)_{i = 1 \dots N}$ and $(v_i)_{i = 1 \dots N}$,
are expressed on the original grid, the differential equations are enforced on
the $N-1$ midpoints $\left(x_{i} + x_{i+1}\right)/2$, the left-hand side
corresponding to finite-difference weights for interpolation.  This leads to
$2N-2$ equations for the $2N$ unknowns.  The remaining $2$ equations are then
provided by the boundary conditions.  Incidentally, this is the second order
approach typically used by many stellar pulsation codes.

In order to set up a higher order scheme, one can consider, for instance,
4-point windows:
\begin{equation}
\begin{array}{rcl}
-\displaystyle \omega \frac{-u_{i-1} + 9 u_i + 9 u_{i+1} - u_{i+2}}{16} &=&
 \displaystyle \frac{v_{i-1} -27 v_i + 27 v_{i+1} - v_{i+2}}{24 \Delta x}, \qquad 2 \leq i \leq N-2, \\
 \displaystyle \omega \frac{-v_{i-1} + 9 v_i + 9 v_{i+1} - v_{i+2}}{16} &=&
 \displaystyle \frac{u_{i-1} -27 u_i + 27 u_{i+1} - u_{i+2}}{24 \Delta x}, \qquad 2 \leq i \leq N-2, \\
-\displaystyle \omega \frac{3 u_1 + 6 u_2 - u_3}{8} &=&
 \displaystyle \frac{-v_1 + v_2}{\Delta x}, \\
 \displaystyle \omega \frac{3 v_1 + 6 v_2 - v_3}{8} &=&
 \displaystyle \frac{-u_1 + u_2}{\Delta x}, \\
-\displaystyle \omega \frac{-u_{N-2} + 6 u_{N-1} + 3 u_N}{8} &=&
 \displaystyle \frac{-v_{N-1} + v_{N}}{\Delta x}, \\
 \displaystyle \omega \frac{-v_{N-2} + 6 v_{N-1} + 3 v_N}{8} &=&
 \displaystyle \frac{-u_{N-1} + u_{N}}{\Delta x}, \\
 u_1 &=& 0,  \qquad  u_N = 0.
\end{array}
\label{eq:four_points}
\end{equation}
where we have assumed a uniform grid when setting up the finite-difference
weights, and have used 3-point windows on the edges to avoid introducing
supplementary bands in the derivation matrix.  Once more, although the unknowns
are expressed on the original $N$ grid points, the difference equations are
enforced on the $N-1$ midpoints, thereby leaving room for the two boundary
conditions.

\begin{figure}[htbp]
\begin{center}
\includegraphics[width=0.8\textwidth]{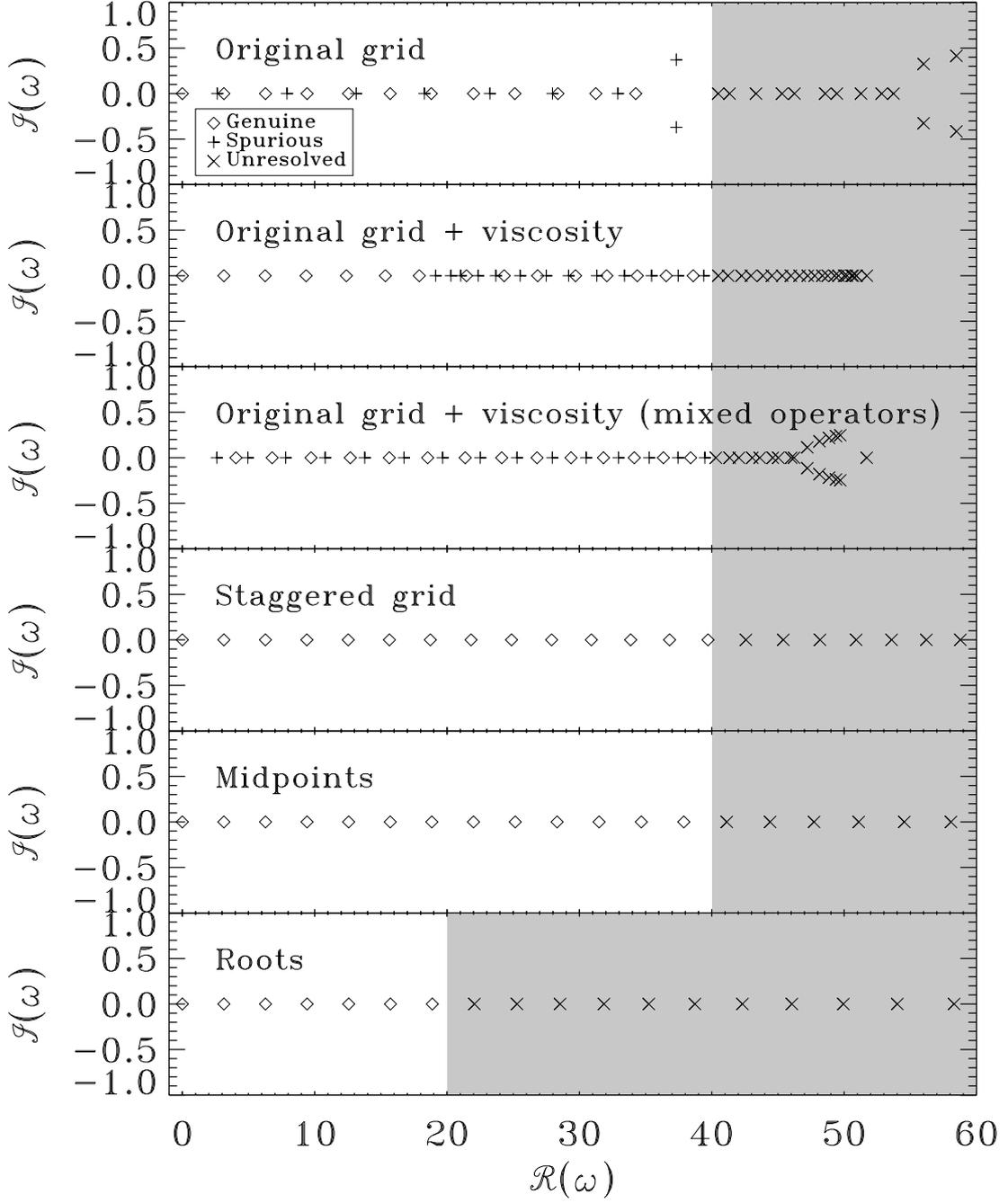}
\end{center}
\caption{Eigenvalues with positive real parts obtained with a QZ algorithm for
various finite-difference schemes.  The labelling of the eigenvalues is based on
a visual inspection of the corresponding eigenmodes.  The shaded grey area
corresponds to solutions which are not fully resolved  -- we required 
approximately 8 points per oscillation period of the genuine solutions (in the
sixth panel, this area is larger simply because we used a non-uniform grid).  In
all cases, the resolution is $N=51$.\label{fig:eigenvalues}}
\end{figure}

The fifth panel of Fig.~\ref{fig:eigenvalues} shows the eigenvalues obtained
using the second scheme with 4-point windows.  This can be contrasted with the
first panel which shows the eigenvalues obtained with the 5-point window scheme,
enforced on the original grid.  As can be seen, the scheme on the alternate grid
has filtered out the spurious eigenvalues, and no complex eigenpairs appear.

\begin{figure}[htbp]
\begin{center}
\includegraphics[width=0.8\textwidth]{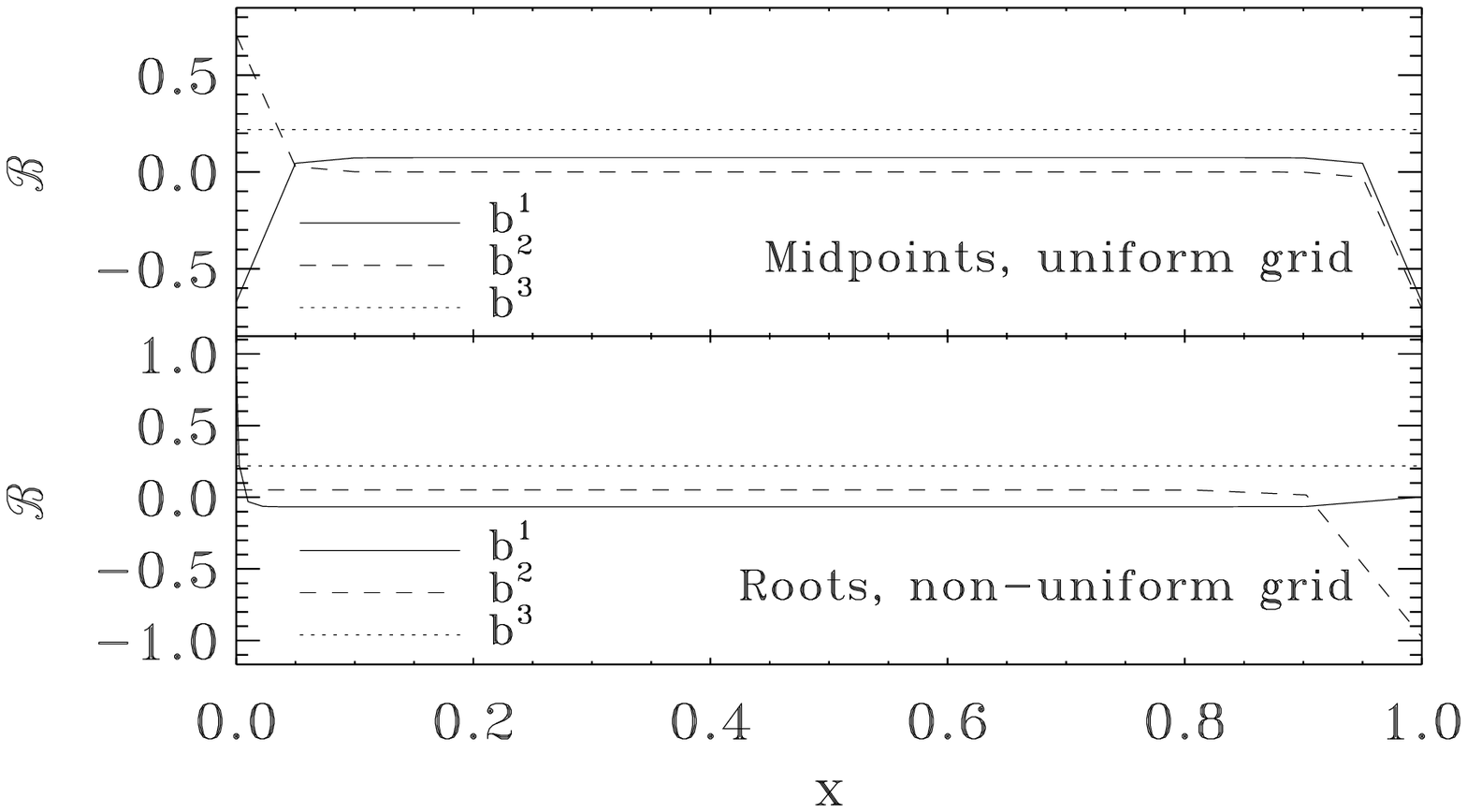}
\end{center}
\caption{Null space basis functions for finite-difference schemes enforced on
alternate grids (midpoints and roots).  The functions $(b^1, b^2 ... b^D)$ are
ordered such that the most oscillatory function comes
first.\label{fig:null_space_bis}}
\end{figure}

We then calculate the null spaces of these new schemes, using the method
described in the previous section.  For the 2-point scheme, the dimension of the
null space is 1 and only contains constant solutions.  The upper panel of
Fig.~\ref{fig:null_space_bis} shows the basis $\mathcal{B}$ for the 4-point
scheme.  In complete contrast to the results shown in Fig.~\ref{fig:null_space},
the basis functions are close to constant in the middle part of the domain, and
only deviate near the edges, where $f'=0$ is not enforced.  This explains the
lack of spurious solutions with mesh drift.

Nonetheless, by making the derivation operator sensitive to oscillatory
solutions, this also leads to introducing interpolation operators which tolerate
such solutions.  One may therefore wonder why the spurious solutions were still
filtered out by using an alternate grid?  A tentative answer to this question is
that in the first case, oscillatory functions are tolerated by the derivation
operators.  These then remain unchanged by the ``interpolation'' operators
(which in this case were simply the identity operator).  In the second case, it
is the interpolation operators which tolerate oscillatory functions. When the
derivation operators are applied to such functions, the oscillatory behaviour is
amplified by an amount roughly proportional to $1/\Delta x$.  Hence, one can
expect the overall system to discard such solutions more easily.

\subsection{Comparison with other methods for suppressing mesh drift}
\label{sect:comparison}

Other solutions have been proposed for solving mesh drift.  For instance,
\citet{Kreiss1978} adds numerical viscosity to the differential system
\citep[see also][]{Press2007}. This can be applied in the following manner to
Eq.~(\ref{eq:example}):
\begin{equation}
\begin{array}{rclrcl}
-\omega u_i &=& \displaystyle \frac{v_{i+1} - v_{i-1}}{2\Delta x} + \epsilon \frac{u_{i-1} -2 u_i + u_{i+1}}{(\Delta x)^2}, \qquad 2 \leq i \leq N-1, \\
 \omega v_i &=& \displaystyle \frac{u_{i+1} - u_{i-1}}{2\Delta x} + \epsilon \frac{v_{i-1} -2 v_i + v_{i+1}}{(\Delta x)^2}, \qquad 2 \leq i \leq N-1, \\
-\omega u_N &=& \displaystyle \frac{v_N - v_{N-1}}{\Delta x}, \qquad
 \omega v_1  =  \displaystyle \frac{u_2 - u_1}{\Delta x}, \\
        u_1 &=& 0,  \qquad   u_N = 0,
\end{array}
\label{eq:numerical_viscosity}
\end{equation}
where $\epsilon$ is a free parameter $\ll 1$.  Although the order of the system
has been increased by $2$, we did not include extra boundary conditions since
this allows a smooth transition between the case $\epsilon = 0$ and $\epsilon
\neq 0$, and is not problematic from a numerical point of view.  By choosing
$\epsilon = 2 \times 10^{-3}$ (for $N=51$), spurious eigenvalues are removed
from the interval $\omega \in [-18;18]$, as is displayed in
Fig.~\ref{fig:eigenvalues} (second panel).  We note that the genuine
eigenfunctions beyond this interval have a somewhat mixed character, as they are
influenced by the neighbouring spurious eigenfunctions, and are not always easy
to identify.  Increasing the value of $\epsilon$ results in a larger interval 
containing only genuine eigenmodes.

\begin{figure}[htbp]
\begin{center}
\includegraphics[width=0.8\textwidth]{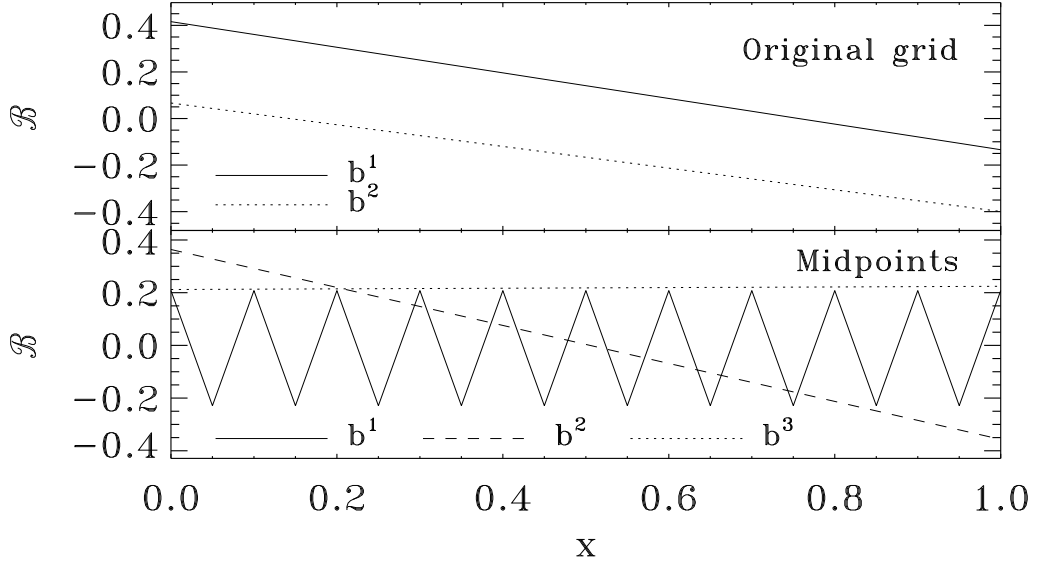}
\end{center}
\caption{Null space basis functions for 2$^\mathrm{nd}$ derivative operators
enforced on the original grid (upper panel) and on the midpoints (lower
panel).  The functions $(b^1, b^2 ... b^D)$ are ordered such that the most
oscillatory function comes first.\label{fig:null_space_2nd_order}}
\end{figure}

The main reason why this method works is because it introduces a second order
derivation operator which is robust to mesh drift when expressed on the original
grid, as opposed to first order derivation operators.  To illustrate this point
more clearly, we set up another somewhat artificial system where the second
order operators are enforced on the midpoints preceding the current grid points,
whereas the other operators are enforced on the original grid:
\begin{equation}
\begin{array}{rclrcl}
-\omega u_i &=& \displaystyle \frac{v_{i+1} - v_{i-1}}{2\Delta x} + \epsilon \frac{u_{i-2} - u_{i-1} - u_i + u_{i+1}}{2(\Delta x)^2}, \qquad 3 \leq i \leq N-1, \\
 \omega v_i &=& \displaystyle \frac{u_{i+1} - u_{i-1}}{2\Delta x} + \epsilon \frac{v_{i-2} - v_{i-1} - v_i + v_{i+1}}{2(\Delta x)^2}, \qquad 3 \leq i \leq N-1, \\
-\omega u_2 &=& \displaystyle \frac{v_3 - v_1}{2\Delta x} + \epsilon \frac{u_1 - 2u_2 + u_3}{(\Delta x)^2}, \\
 \omega v_2 &=& \displaystyle \frac{u_3 - u_1}{2\Delta x} + \epsilon \frac{v_1 - 2v_2 + v_3}{(\Delta x)^2}, \\
-\omega u_N &=& \displaystyle \frac{v_N - v_{N-1}}{\Delta x} + \epsilon \frac{u_{N-2} - 2 u_{N-1} + u_N}{(\Delta x)^2}, \\
 \omega v_1 &=&  \displaystyle \frac{u_2 - u_1}{\Delta x}, \qquad
        u_1  = 0,  \qquad   u_N = 0.
\end{array}
\end{equation}
The associated eigenvalues are shown in Fig.~\ref{fig:eigenvalues} (third
panel), where $\epsilon = 2 \times 10^{-3}$.  This time no intervals containing
only genuine eigenmodes appear.  If $\epsilon$ is further increased, the genuine
and spurious eigenmodes start to interact and form complex eigenpairs.  One can
also calculate the basis functions of the null spaces associated with these two
2$^\mathrm{nd}$  derivative operators by applying the same procedure as in
Section~\ref{sect:characterising}.  The results are shown in
Figure~\ref{fig:null_space_2nd_order}.  As expected, the first operator (upper
panel) only has two basis functions, both of which are affine functions (which
are true solutions to the mathematical problem).  The second operator (lower
panel) has a third basis function which is highly oscillatory.  If we reapply
the same intuitive reasoning as earlier, we can expect the second derivative of
an oscillatory function to vanish near the inner midpoints of a given window of
grid points, hence the reason for this supplementary, oscillatory basis
function.

An important drawback with this method is that it introduces a supplementary
free parameter, $\epsilon$, which needs to be carefully adjusted.  On the one
hand, if $\epsilon$ is too small, then the system will not be robust to mesh
drift.  On the other, a large value of $\epsilon$ alters the original system and
reduces the overall accuracy.  Another difficulty with this approach is that it
is not always straightforward where and how to introduce the 2$^\mathrm{nd}$
derivative operators.  For instance, swapping the two terms proportional to
$\epsilon$ in Eq.~(\ref{eq:numerical_viscosity}) is highly detrimental to the
solutions and does not solve mesh drift.  The reason for this behaviour is not
obvious and is beyond the scope of this paper.

Another solution used for suppressing mesh drift consists in applying a
staggered grid approach.  For instance, the variable $u$ could be defined on the
original grid, and $v$ on the midpoints.  This would lead to the following
difference equations:
\begin{equation}
\begin{array}{rcl}
-\displaystyle \omega u_i &=&
 \displaystyle \frac{v_i - v_{i-1}}{\Delta x}, \qquad 2 \leq i \leq N-1, \\
 \displaystyle \omega v_i &=& \displaystyle \frac{u_{i+1} - u_i}{\Delta x}, \qquad 1 \leq i \leq N-1, \\
 u_1 &=& 0,  \qquad  u_N = 0,
\end{array}
\label{eq:staggered_grid}
\end{equation}
where we have assumed a uniform grid.  The above system contains $2N-3$
difference equations and $2$ boundary conditions for $2N-1$ unknowns.  By
applying a QZ algorithm, we see that the spurious eigenvalues have once more
been filtered out (Figure~\ref{fig:eigenvalues}, fourth panel).  The same
mechanism which filtered out mesh drift in the alternate grid approach also
applies to the staggered grid approach.  However, the latter approach also comes
with some disadvantages.  Firstly, there's the inconvenience of having different
variables defined on different grids. Furthermore, if we had introduced, say,
$3$ variables in a 1D case, it is not entirely clear which variables should be
defined on what grids.  If one defines the two first variables on the original
grid and the third one on the midpoints, they may be confronted with a
differential equation which relates the two first variables and which
consequently does not block mesh drift, since these variables are on the same
grid (this may not be a problem if the remaining equations are able to suppress
mesh drift).  Another solution could consist in defining one of the variables on
the original grid, a second one on the one-third points, and the third one on
the two-thirds points.  However, such an approach might be less effective at
blocking mesh drift and certainly lacks the simplicity of using the alternate
grid approach.  Another issue concerns the boundary conditions. If the boundary
conditions in Eq.~(\ref{eq:example}) had been, for example, $u(0) + v(0) = 0$
and $u(1) - v(1) = 0$, then one needs to deal with the values of both functions
at both endpoints.  However, the variable $v$ is defined on a grid which does
not include the endpoints.  As a result, some ad-hoc treatment of the boundary
conditions is required.  Finally, as will be made clear in the following
section, the staggered grid approach does not always lend itself to
superconvergence.

Rhie-Chow interpolation has also been frequently used to suppress mesh drift in
fluid dynamics problems \citep{Rhie1983}.  This approach involves interpolating
the velocity, using the momentum equation in which the pressure gradient has
been replaced with a more stable expression only involving adjacent grid
points.  As a result, adjacent grid points become coupled, thereby suppressing
mesh drift.  This approach has the advantage of keeping all of the
variables on the original grid, as opposed to the staggered grid approach. 
However, its main drawback is that it is specific to the momentum equation and
doesn't seem to have been extended into a general mathematical approach.
Furthermore, it was originally developed using second order finite-differences,
and it is not clear whether it extends to higher orders.

\section{Superconvergence}
\label{sect:superconvergence}

One last point of interest concerns the order of the finite-difference
operators. In Eq.~(\ref{eq:two_points}), the overall system has a second order
accuracy, whether or not the underlying grid is uniform.  In
Eq.~(\ref{eq:four_points}), the overall accuracy is of third order, in spite of
having increased the window size from two to four.  This is because, the
derivation operators on the endpoints only locally achieve a second order
accuracy.  Furthermore, if the underlying grid had been non-uniform, then the
difference equations in the middle part of the domain would have been of third
rather than fourth order accuracy, when enforced on the midpoints.  This is
because, in general, finite differences calculated over a window of $n$ grid
points are accurate to order $n-d$, where $d$ is the order of the derivative. 
However, the scheme used in Eq.~(\ref{eq:two_points}) is an exception to this
rule.  Furthermore, \citet{Sadiq2011} recently pointed out general conditions
where the accuracy is boosted by one or several orders, a
situation which they called ``superconvergence''.  This therefore raises the
question as to whether such conditions can systematically be satisfied by
carefully choosing the alternate grid on which to enforce the finite-difference
scheme.  In what follows, we show that the answer to this question is yes
for an increase in accuracy by one order.  We provide our own
derivation of the condition for superconvergence and show that it is compatible
with removing mesh drift.  Our derivation is somewhat simpler than what is
presented in \citet{Sadiq2011} and leads to a formulation which is easily
implemented numerically.

Let $y$ be a function for which we're seeking to approximate the $d$-th
derivative at a specified point, $z$, using finite differences.  Let  $x_1, x_2
\dots x_n$ be a set of $n$ distinct grid points for which the function $y$ is
known.  A Taylor expansion centred on $z$ of the function $y$ yields the
following expression at each grid point $x_i$:
\begin{equation}
y(x_i) = y(z) + y'(z) (x_i-z) + \dots + y^{(n)}(z) \frac{(x_i-z)^n}{n!}
       + \mathcal{O}\left(\left(x_i-z\right)^{n+1}\right).
\end{equation}
Finite-difference coefficients $a_1^d(z), a_2^d(z) \dots a_n^d(z)$ are
then calculated so as to satisfy:
\begin{equation}
\label{eq:FD}
\sum_{i=1}^{n} a_i^d(z) y(x_i) = y^{(d)}(z) + R,
\end{equation}
where
\begin{equation}
\label{eq:R_def}
R = \sum_{i=1}^{n} a_i^d(z) \left\{ y^{(n)}(z) \frac{(x_i-z)^n}{n!}
       + \mathcal{O}\left(\left(x_i-z\right)^{n+1}\right) \right\}.
\end{equation}
This is achieved by having the coefficients satisfy the following
relationships:
\begin{equation}
\sum_{i=1}^{n} a_i^d(z) \frac{(x_i-z)^j}{j!} = \delta_{jd}, \qquad 0 \leq j \leq n-1,
\end{equation}
where $\delta_{jd}$ is Kronecker's delta symbol.  Despite appearances, $R$ is
actually of order $\mathcal{O}\left(\delta x^{n-d}\right)$ in most cases, where
$\delta x$ is representative of the separations between the $x_i$.

The goal is then to find $z$ values for which the first part of $R$ cancels out:
\begin{equation}
\label{eq:problem}
0 = \sum_{i=1}^{n} a_i^d(z) y^{(n)}(z) \frac{(x_i-z)^n}{n!}.
\end{equation}
Finding such values would guarantee that $R$ is a least of order
$\mathcal{O}\left(\delta x^{n-d+1}\right)$.

It turns out that the right hand side of Eq.~(\ref{eq:problem}) is a product
of a polynomial of degree $n-d$ and of the $n$-th derivative of $y$:
\begin{equation}
\label{eq:solution}
\sum_{i=1}^{n} a_i^d(z) y^{(n)}(z) \frac{(x_i-z)^n}{n!} = -\frac{y^{(n)}(z)}{n!} Q^{(d)}(z),
\end{equation}
where 
\begin{equation}
Q(x) = \prod_{i=1}^{n} (x-x_i).
\end{equation}
It is straightforward to see that the polynomial $Q^{(d)}$ has $n-d$ distinct
zeros, which may be obtained by a standard root solver.  We also note, in passing,
that Eq.~(\ref{eq:solution}) agrees with the first term of the Taylor expansion
of the finite-difference errors given in \citet{Bowen2005}.

We start by calculating $R$ for the function $y_0(x) = x^n$.  Based on
Eq.~(\ref{eq:FD}), one has:
\begin{equation}
R = - y_0^{(d)}(z) + \sum_{i=1}^{n} a_i^d(z) x_i^n.
\end{equation}
According to \citet{Fornberg1988}, the coefficients $a_i^d(z)$ are given by:
\begin{equation}
a_i^d(z) = L_i^{(d)}(z),
\end{equation}
where $L_i(z)$ is Lagrange's interpolation polynomial defined such that
$L_i(x_j) = \delta_{ij}$.  Let us define a polynomial $P$ as follows:
\begin{equation}
P(x) = -x^n +  \sum_{i=1}^{n} L_i(x) x_i^n.
\end{equation}
It turns out that $P^{(d)}(z) = R$.  Furthermore:
\begin{equation}
P(x_i) = -x_i^n + x_i^n = 0, \qquad 1 \leq i \leq n.
\end{equation}
However, $P$ is not identically zero but is of degree $n$, with a leading
coefficient of $-1$.  From this, we deduce that $P = -Q$.

Based on its definition, Eq.~(\ref{eq:R_def}), $R$ is also given by:
\begin{equation}
R = \sum_{i=1}^{n} a_i^d(z) y_0^{(n)}(z) \frac{(x_i-z)^n}{n!}
  = \sum_{i=1}^{n} a_i^d(z) (x_i-z)^n,
\end{equation}
where we have used $y_0^{(n)}(z) = n!$.  Hence, one obtains the
following relationship:
\begin{equation}
\sum_{i=1}^{n} a_i^d(z) (x_i-z)^n = -Q^{(d)}(z).
\end{equation}
When inserted into the right hand side of Eq.~(\ref{eq:problem}),
this gives the desired result, \textit{i.e.} Eq.~(\ref{eq:solution}).

\begin{figure}
\begin{minipage}{0.6\textwidth}
\includegraphics[width=\textwidth]{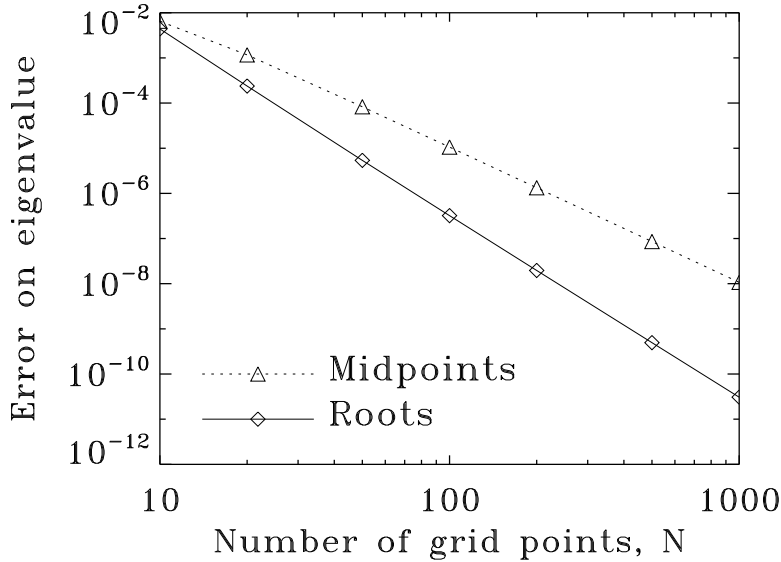}
\end{minipage} \hfill
\begin{minipage}{0.37\textwidth}
\caption{Error on the $\omega=\pi$ eigenvalue of Eq.~(\ref{eq:example}) as a
function of the number of grid points, $N$.  Using the roots of $Q'$ increases
the rate of convergence from $N^{-3}$ to $N^{-4}$. \label{fig:errors}}
\end{minipage}
\end{figure}

If we now consider the case $d=1$, the $n-1$ roots of $Q'$ are interspersed with
the $n$ original grid points in the window.  Enforcing the finite differences
on, say, the middle root (assuming $n$ is even) therefore achieves the same
result as using a midpoint, given that it stays far away from the grid points,
and can consequently be expected to remove mesh drift. To verify this, we
calculated the null space of a 4-point derivation operator calculated for the
non-uniform grid based on Eq.~(\ref{eq:grid_non_uniform}) and enforced on the
middle root of $Q'$, except near the endpoints where the window size is 3, in
which case we choose the root nearest the endpoints.  The basis functions of the
null space are presented in the lower panel of Fig.~\ref{fig:null_space_bis} and
display no oscillatory behaviour.  If we return to our initial problem
(Eq.~(\ref{eq:example})), spurious eigenvalues are once more filtered out
(Fig.~\ref{fig:eigenvalues}, last panel).  Finally, in Fig.~\ref{fig:errors}, we
show the rate of convergence of the $\omega=\pi$ eigenvalue using the same
non-uniform grid, and enforcing the finite differences on the midpoints and on
the middle roots (dotted and continuous line, respectively).  As expected, the
error scales as $N^{-3}$ when working with the midpoints and as $N^{-4}$ when
working with the middle roots.

At this point, it is fairly straightforward to see why the staggered grids
approach does not always lend itself to superconvergence.  Indeed, if we take
our initial problem as an example, to achieve superconvergence the variable $v$
needs to be defined on the root points of the grid associated with $u$,
\textit{and vice-versa}.  This leads to two relationships between the two grids
which are not necessarily compatible.  If, for instance we consider two-point
windows for simplicity, then the root points will coincide with the midpoints.
To then achieve superconvergence, the midpoints of the midpoints of $u$'s grid
need to coincide with the original grid points.  This is only possible if the
original grid is uniform.

\section{Application to stellar pulsations}
\label{sect:application}

In practical situations, one will want to solve differential systems which are
far more complicated than Eq.~(\ref{eq:example}) and typically involve
non-constant coefficients.  As a result, to apply the above method, it will be
necessary to interpolate these coefficients.  This can easily be done, to the
same order of accuracy, by simply reusing the same interpolation weights that
were used on the variables.  Also, in some situations, the underlying grid can
be quite dense in areas where either the coefficients or the expected solutions
vary rapidly.  This can lead to poor results when calculating the roots of
$Q'$.  A simple remedy consists in applying an affine transformation to each
window of grid points, in order to spread them out and recenter them around $0$,
before finding the roots of $Q'$.  The inverse transformation then needs to be
applied to the resultant roots, in order to obtain them with respect to the
original grid. Although, both of these may represent extra complications
compared to applying finite differences on the original grid points, the gain in
numerical stability and accuracy makes this approach worthwhile.

We now demonstrate that the above approach can successfully be applied to
stellar pulsations.  In what follows we will apply a fourth order approach using
four point windows and superconvergence.  Based on the preceding results, such
an approach reaches fourth order accuracy.  In a first example, we focus on an
acoustic mode in a polytropic stellar model and show this approach is both
accurate and robust, even for a highly uneven grid, where Richardson
extrapolation fails. In the second example, we apply this method to calculating
oscillations in a realistic stellar model, namely Model S
\citep{Christensen-Dalsgaard1996} and show that it produces results in agreement
with previous approaches and filters out spurious modes.  Then we briefly
describe how this approach is applied to 2D calculations of pulsation modes in
rapidly rotating stars.

\subsection{Richardson extrapolation on uneven grids}
\label{sect:Richardson}

As a first example, we focus on the $(n,\,\ell) = (4,1)$ acoustic mode in an
$N_{\mathrm{p}}=3$ polytropic stellar model, were $n$ is the radial order,
$\ell$ the harmonic degree, and $N_{\mathrm{p}}$ the polytropic index.  We
compare two approaches: the fourth order scheme described above, and a second
order scheme supplemented with Richardson extrapolation.  Two types of grids are
considered: a uniform grid, and an uneven grid in which consecutive separations
alternate between a large and a small value.  The ratio of the small value to
the large value has been set to $0.1$.

\begin{figure}
\begin{minipage}{0.6\textwidth}
\includegraphics[width=\textwidth]{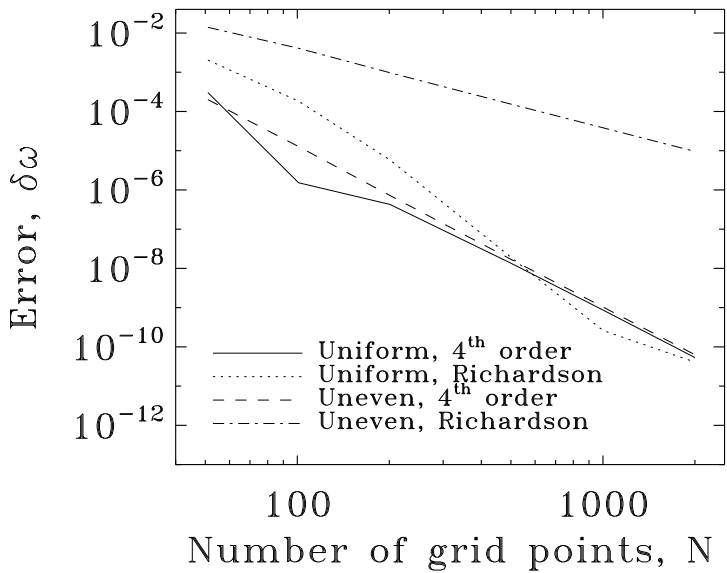}
\end{minipage} \hfill
\begin{minipage}{0.37\textwidth}
\caption{Error on the frequency as a function of the grid resolution for a
2$^{\mathrm{nd}}$ order approach combined with Richardson extrapolation and a
4$^{\mathrm{th}}$ order approach, applied to uniform and uneven grids.
\label{fig:Richardson}}
\end{minipage}
\end{figure}

Figure~\ref{fig:Richardson} then shows the error on the frequency, as a function
of grid resolution, for both types of grids and finite-difference schemes.  A
fourth order calculation using a uniform grid with $5001$ points has been used
to calculate the reference frequency used to evaluate the error (similar
reference values are also obtained using Richardson extrapolation on the same
grid, or using a spectral approach based on Chebyshev polynomials).  When
applying Richardson extrapolation, two calculations are carried out: one on the
full grid, and one on a reduced version of the grid in which every other point
is dropped out.  Hence, the reduced version of the uneven grid turns out to be
uniform, given that the grid point separations alternate between a large a small
value.  As can be seen in the figure, Richardson leads to fourth order accuracy
for the uniform grid but not the uneven grid.  Indeed, the second order errors
have not been removed in the latter case.  In contrast, the fourth order scheme
yields fourth order accuracy, regardless of the grid.  Of course, one must bear
in mind that this situation is somewhat contrived and that the underlying grids
used in most stellar pulsation calculations are close to uniform over short
intervals, even if they are highly uneven over the entire domain.  Hence, in
most practical situations, Richardson extrapolation will improve results and
lead to near fourth order accuracy.

\subsection{Model S}

\begin{figure}
\begin{minipage}{0.6\textwidth}
\includegraphics[width=\textwidth]{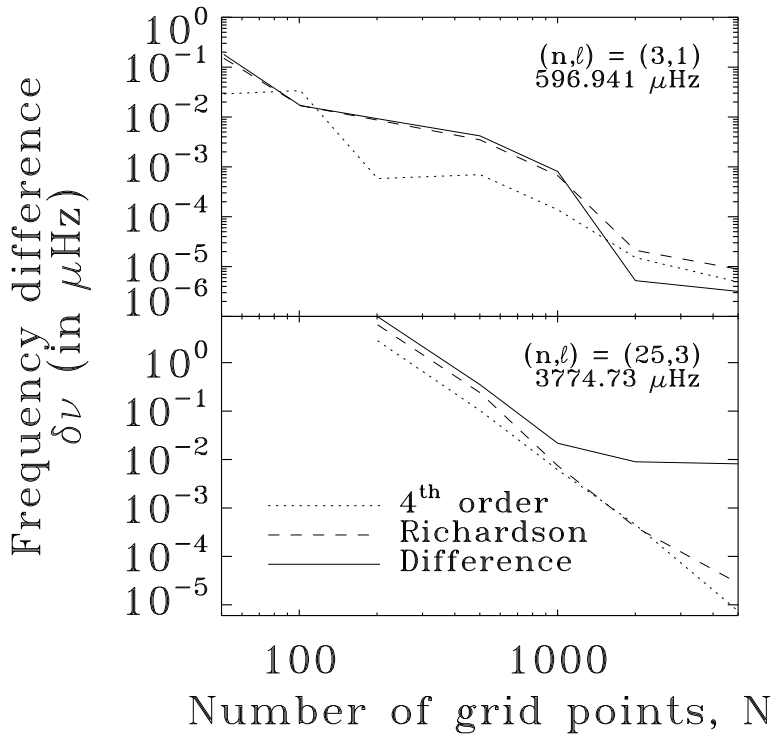}
\end{minipage} \hfill
\begin{minipage}{0.37\textwidth}
\caption{Error on the frequencies of two modes as a function of the grid
resolution for a 2$^{\mathrm{nd}}$ order approach combined with Richardson
extrapolation (implemented in the ADIPLS code) and a 4$^{\mathrm{th}}$ order
approach.  The radial order, harmonic degree, and frequency of each mode
is indicated in the upper right corner of each panel.\label{fig:ModelS}}
\end{minipage}
\end{figure}

In order to test the above scheme in a more realistic situation, we apply it to
calculating stellar pulsations in Model S, a reference model representative of
the current Sun \citep{Christensen-Dalsgaard1996}.  This model is interpolated
onto grids of different resolutions ($N=51$, 101, 201, 501, $\dots$ 10001) using
the program \texttt{redistrb.d} which comes with the ADIPLS package.  The option
appropriate for p-modes was selected thereby causing the underling grid to be
dense near the surface where the sound velocity, and hence the wavelength, is
small.  A simple surface boundary condition $\delta p = 0$ was used when
calculating the pulsation modes.  Figure~\ref{fig:ModelS} shows the errors on
the frequencies of two modes as a function of the grid resolution.  The second
order calculation supplemented with Richardson extrapolation was carried out
using the ADIPLS code \citep{Christensen-Dalsgaard2008}.  The upper and lower
panels correspond to low and high frequency acoustic modes, respectively.  For
each method, the errors on the frequencies were calculated using the frequency
calculation at $10001$ points, for that particular method, as a reference
value.  These errors are shown as dotted and dashed lines for the fourth order
method, and the second order method with Richardson extrapolation,
respectively.  Finally, the solid line indicates the frequency differences
between both methods.  For the low frequency mode, the differences between both
methods is very small and comparable to the errors on the frequencies. We note,
however, that the general slope of the errors is closer to $N^{-2}$ rather than
$N^{-4}$, regardless of the method.  A possible reason for this may be the way
the model is interpolated onto the different grids.  The situation is different
for the high frequency mode.  This time the slope of the error is close to
$N^{-4}$, as expected theoretically.  However, as shown by the solid line, the
two methods converge to values approximately $8$ nHz apart, which is
substantially larger than what was obtained for the low frequency mode. 
Nonetheless, this remains smaller than the difference between the variational
and Richardson frequencies (approximately $58$ nHz).  Hence, these two methods
agree to a high degree of accuracy, thereby validating the fourth order approach
in a realistic situation. 

\begin{figure}
\begin{center}
\includegraphics[width=0.8\textwidth]{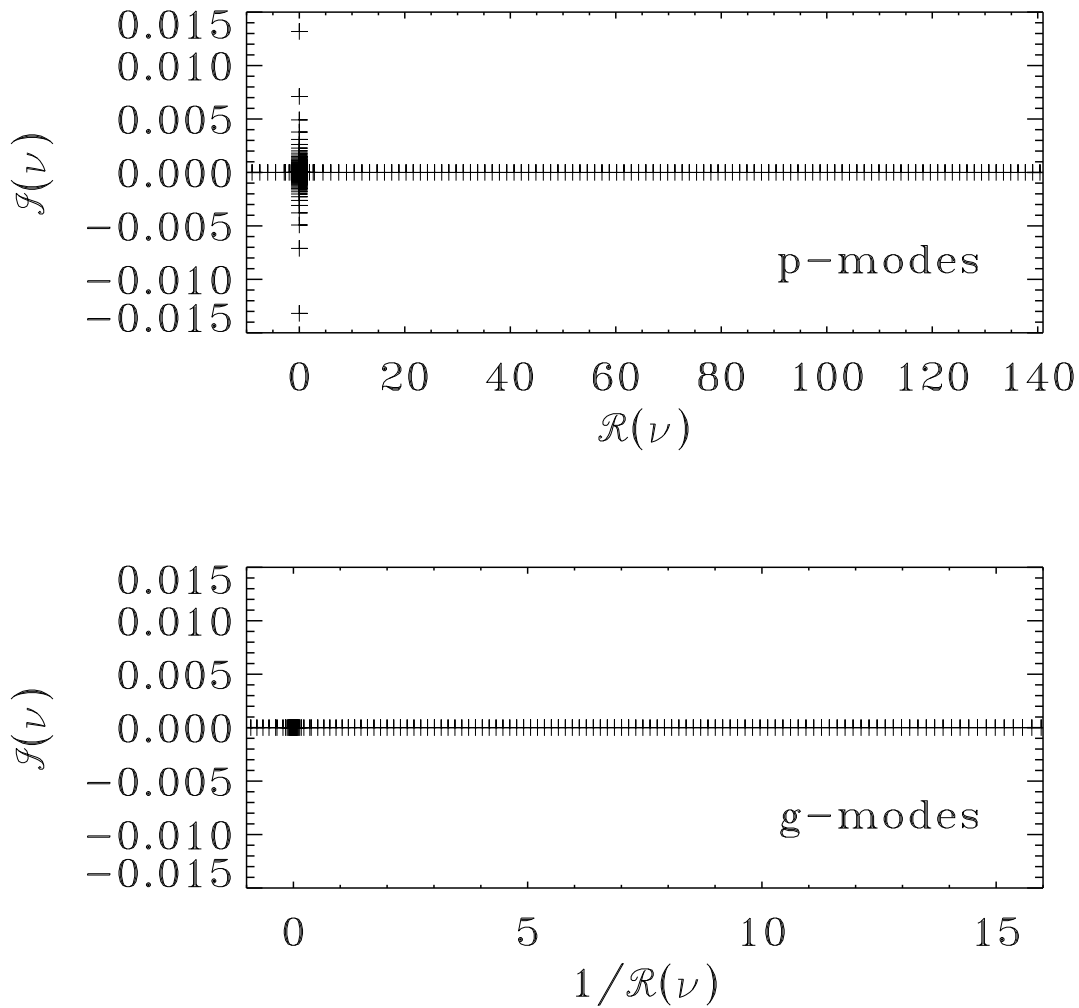}
\end{center}
\caption{Frequency spectrum of $\ell=1$ modes calculated in Model S
(interpolated onto 1001 grid points) using a QZ algorithm.  Spurious modes only
appear at the outskirts of the spectrum where the numerical resolution is
insufficient to resolve the modes. \label{fig:ModelS_spectrum}}
\end{figure}

A second test consisted in numerically calculating an entire $\ell=1$ pulsation
spectrum for the $N=1001$ model, using a QZ algorithm. The QZ algorithm
is an iterative procedure for finding \textit{all} of the solutions of the
generalised eigenvalue problem $Ax = \lambda B x$ where $A$ and $B$ are general
square matrices \citep{Moler1973}. The implementation we used comes from the
LAPACK\footnote{We specifically used the DGGEV routine, which is based on the
DHGEQZ routine.  LAPACK is available at \url{http://www.netlib.org/lapack/}.}
linear algebra library. Figure~\ref{fig:ModelS_spectrum} shows parts of
the frequency spectrum which include the first hundred p and g modes.  A
mixture of both real and complex eigenvalues appear.  Given that the pulsation
calculations are based on the adiabatic approximation, the complex eigenvalues
can only be due to numerical artifact, and must be considered as spurious. 
However, these solutions are located at the outskirts of the frequency
spectrum, \textit{i.e.} where modes are unresolved due to insufficient
grid resolution. In contrast, an inspection of the modes with $-100 \leq n \leq
100$ did not reveal spurious modes, complex eigenvalues, or mesh
drift.  We therefore conclude that this method has successfully removed the
problems related to mesh drift.

\subsection{2D calculations in rapidly rotating stars}

\begin{figure}
\begin{minipage}{0.6\textwidth}
\includegraphics[width=\textwidth]{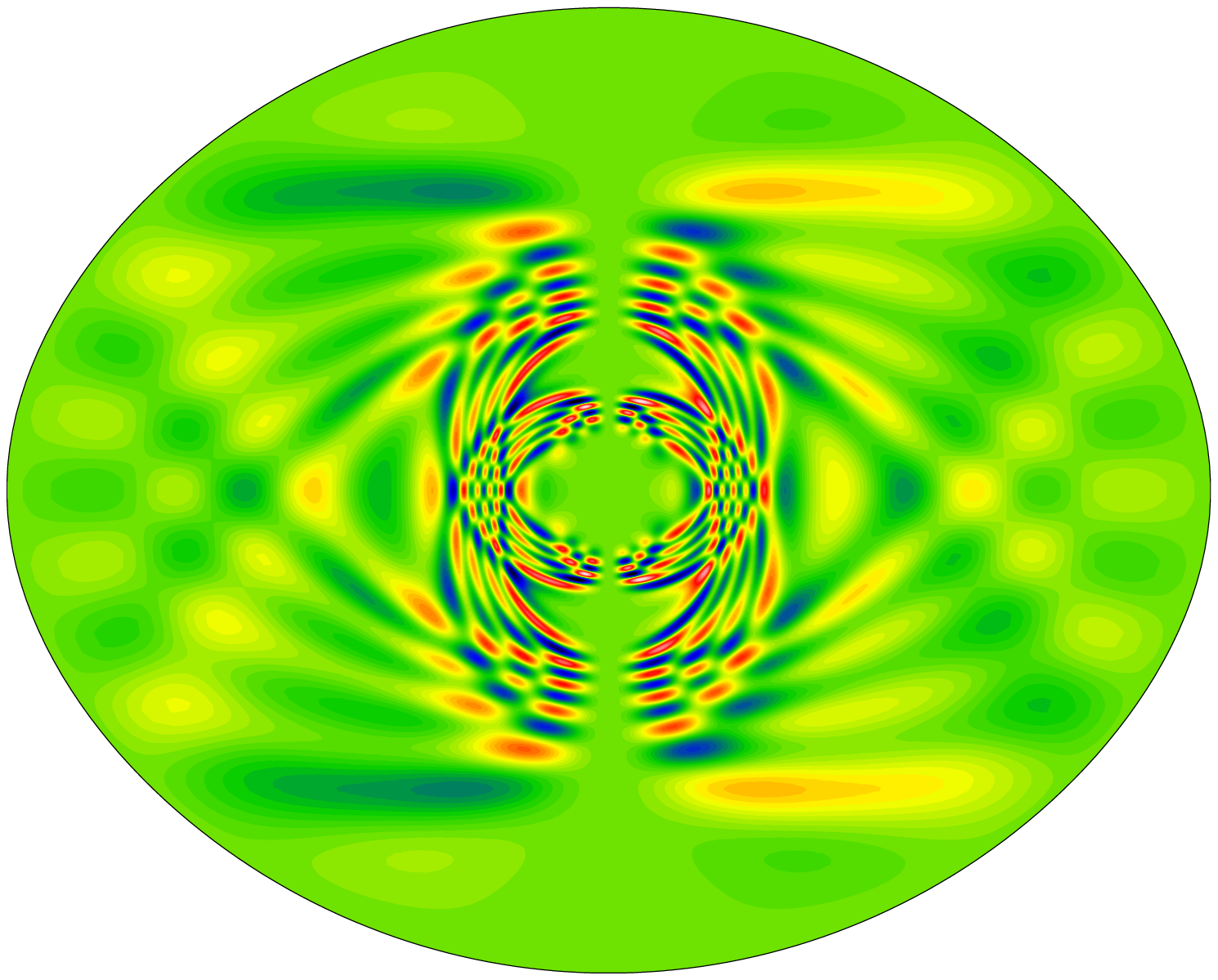}
\end{minipage} \hfill
\begin{minipage}{0.37\textwidth}
\caption{Meridional cross-section of a pulsation mode in a stellar model
rotating at $70\,\%$ of the critical break-up rotation rate. This mode was
calculated using the fourth order scheme with superconvergence in the radial
direction and a spectral method based on spherical harmonics in the angular
direction.  The resolution is 1601 radial grid points and 40 spherical
harmonics.\label{fig:field}}
\end{minipage}
\end{figure}

The fourth order approach has also been successfully applied to calculating
pulsation modes in rapidly-rotating centrifugally-distorted stars, a far more
complicated eigenvalue problem involving 2D partial differential equations. 
Indeed, it has been used for the radial discretisation in the TOP code and
represents a definite improvement over the unstable scheme previously used in
\citet{Reese2009}.  Based on this improvement, \citet{Burke2011} obtained stable
numerical behaviour when calculating the pulsations of centrifugally-deformed
models from the ASTEC code \citep{Christensen-Dalsgaard2008b}, and
\citet{Reese2013} were able to remove spurious modes from the pulsation spectra
of models based on the Self-Consistent Field (SCF) Method \citep{MacGregor2007},
which is important when scanning an entire spectrum.  As an example,
Figure~\ref{fig:field} shows the meridional cross-section of a rosette mode
\citep{Ballot2012} in an SCF stellar model, using this approach.

\section{Conclusion}

In this paper, we have explored different ways of setting up higher order finite
difference schemes for calculating stellar pulsations.  This lead us to
investigating mesh drift and related problems, such as spurious modes.  After
having illustrated and characterised mesh drift, we proposed a simple remedy
which furthermore lends itself to ``superconvergence'', a way of boosting the
order of accuracy of finite differences by one.  We then applied this method to
1D and 2D stellar pulsation calculations, and showed that it was accurate,
flexible with respect to underlying grid, and able to remove to mesh drift. 
Finally, we should emphasise that although this approach was specifically
developed for calculating stellar pulsations, it is easily applicable and
beneficial to many other physical or mathematical problems.

\begin{acknowledgements}
DRR thanks the referee, Michel Rieutord, and François Lignières for
useful comments and suggestions which have improved and clarified the
manuscript.  DRR is financially supported through a postdoctoral fellowship from
the ``Subside fédéral pour la recherche 2011'', University of Liège, and was
previously supported by the CNES (``Centre National d'Etudes Spatiales''), both
of which are gratefully acknowledged.
\end{acknowledgements}

\bibliographystyle{aa}
\bibliography{biblio}

\end{document}